\documentclass[preprint,aps,nofootinbib]{revtex4}
\usepackage{mathrsfs}
\usepackage{graphicx}
\usepackage{amsmath}
\usepackage{amsfonts}
\usepackage{amssymb}
\usepackage{color}
\newcommand{\fig}[1]{Fig. \ref{#1}}
\newcommand{\bean}{\begin{eqnarray}}
\newcommand{\eean}{\end{eqnarray}}

\newcommand{\eqs}[1]{Eqs. (\ref{#1})}
\newcommand{\eq}[1]{Eq. (\ref{#1})}
\newcommand{\meq}[1]{(\ref{#1})}
\newcommand{\ppa}[2]{\left(\frac{\partial}{\partial #1}\right)^{#2}}

\newcommand{\ecm}{E_{c.m.}}
\newcommand{\rh}{r\!_H}

\begin{document}
\title{Particle collisions near the cosmological horizon of a Reissner-Nordstr\"om de Sitter black hole}
\author{Changchun Zhong}
\email{cczhong@mail.bnu.edn.cn} \affiliation{Department of Physics, Beijing
Normal University, Beijing 100875 P.R. China}
\author{Sijie Gao}
\email{sijie@bnu.edn.cn} \affiliation{Department of Physics, Beijing
Normal University, Beijing 100875 P.R. China}

\begin{abstract}
It has recently been shown that  black holes can act as
particle accelerators and two particles can collide with arbitrarily high center-of-mass (CM) energy
under certain critical conditions. In this paper, we investigate particle collisions outside a Reissner-Nordstr\"om de Sitter (RN-dS) black hole. We find that infinite CM energy can be produced near the cosmological  horizon for generic spacetime configurations. Remarkably, such infinite CM energy does not require the black hole to be extremal, in contrast to spacetimes in the absence of cosmological constants. However, since the charge of an astrophysical body is negligible, the required charge to mass ratio of the particle is extremely higher than that of any elementary  particle.
\end{abstract}

\maketitle

\section{Introduction}
 Banados, Silk and West (BSW)\cite{MJS} showed that Kerr black holes can serve as particle accelerators and infinite center-of-mass energies can in principle arise. The BSW mechanism was soon extended to different black hole solutions
\cite{TT}-\cite{mp}
. These works suggest that the following features are required for the divergence of the CM energy: (1) The collision must occur arbitrarily close to the horizon. (2) One of the particles  possesses a critical value of angular momentum or charge; (3) The black hole must be extremal. It has been pointed out \cite{TT,comment}that condition (3) can not be fulfilled due to the theoretical upper limit on the spin parameter of black hole. So it would be more meaningful to look for infinite CM energies around a non-extremal black hole. Recently, we have proven\cite{gsccz} that infinite CM energies cannot be created outside a non-extremal Kerr black hole. However, Wei et.al. \cite{yangjie} pointed out that for a Kerr de-Sitter black hole, two particles can collide with arbitrarily high CM energy without imposing the extremal condition. This result indicates that if cosmological constant is taken into account, condition (3) may be released.
Since the well-known $\Lambda CDM$ model fits remarkably well with the current cosmological observations, it is worthwhile to further study the accelerating effect for spacetimes with a positive cosmological constant.

In this paper, we investigate particle collisions near a charged black hole in an asymptotically de-Sitter spacetime, i.e., Reissner-Nordstr\"om de-Sitter black hole. Previously, Zaslavskii \cite{OBZ} has studied the radial motion of charged particles in a Reissner-Nordstr\"om (RN) background and found the result is similar to that in a Kerr background. In particular, infinite CM energies can only be attained when the black hole is extremal, i.e., the black hole possesses the maximum charge $Q=M$. However, the situation will be different if a positive cosmological constant $\Lambda$ is introduced. In an asymptotically de Sitter black hole, there exists a cosmological horizon which is located at the radius of order $\Lambda^{-1/2}$. By calculating the radial motion of charged particles, we find that infinite CM energies can be obtained at the cosmological horizon of a generic RN-dS black hole. In this case, the black hole need not  be extremal. However, a critical charge is required for one of the particles. By numerical estimation, we find that the charge to mass ratio of the particle is much higher than that of an electron. Thus, the infinite CM energy is not realizable in the real world.

This paper is organized as follows. In section \ref{s1}, we discuss the horizons of RN-dS space-times. In section \ref{s2}, we calculate
the CM energy of two particles near the cosmological horizon of RN-dS black hole. In section \ref{s3}, we perform some numerical calculations based on the data of astrophysical observations.  Finally, conclusions are made in section \ref{conc}.

\section{RN-dS Black Hole Horizons}\label{s1}
\begin{figure}[!htbp]
\centering
\begin{picture}(240,180)
\put(0,57){\vector(1,0){270}}
\put(15,30){\vector(0,1){140}}
\put(0,175){$\alpha(r)$}
\put(260,60){$r$}
\put(130,100){$\alpha=1-\frac{2M}{r}+\frac{Q^2}{r^2}-\frac{r^2}{l^2}$}
\put(60,135){$\alpha=1-\frac{2M}{r}+\frac{Q^2}{r^2}$}
\put(17,50){$r_1$}\put(45,50){$r_2$}\put(240,62){$r_3$}
\includegraphics{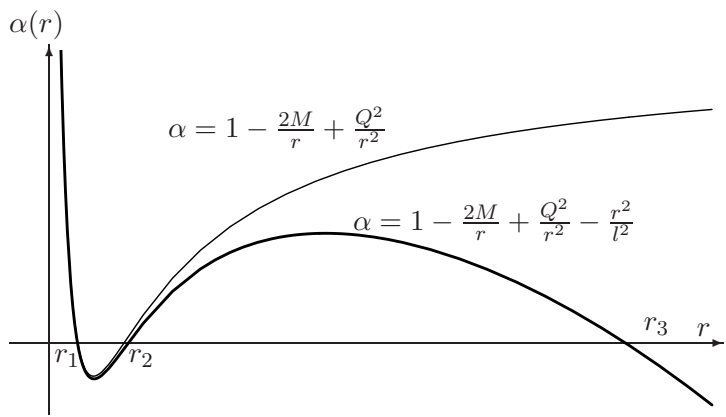}
\end{picture}
\caption{The function $\alpha(r)$ for the RN and RN-dS black holes.\label{fg1}}
\end{figure}

The RN-dS black hole is described by the following metric
\begin{equation}\label{ds2}
ds^2=-\alpha dt^2+\frac{dr^2}{\alpha}+r^2d\theta^2+r^2\sin^2\theta
d\varphi^2,
\end{equation}
where $\alpha=1-\frac{2M}{r}+\frac{Q^2}{r^2}-\frac{r^2}{l^2}$ and
$l=\sqrt{\frac{3}{\Lambda}}$. The roots of $\alpha=0$ give rise to the location of horizons. We denote the three roots by $r_1, r_2, r_3$. $r_1$ and $r_2$ correspond to the inner and outer horizons of a RN black hole, while $r_3$ is called the cosmological horizon. Due to the smallness of $\Lambda$, $r_3$ can be approximated by $l$ and is much larger than $r_1$ and $r_2$. \fig{fg1} depicts the function $\alpha(r)$ in the RN and RN-dS cases. For $l \gg M$, the radius of cosmological horizon $r_3$ is much larger than the radius of  the two black hole horizons $r_1$ and $r_2$.

\begin{figure}[!htbp]
\centering
\begin{picture}(400,100)
\put(0,22){\vector(1,0){140}}
\put(7,10){\vector(0,1){80}}
\put(-5,90){$\alpha$}
\put(130,26){$r$}
\put(16,15){$r_{1,2}$}
\put(120,15){$r_3$}

\put(140,50){\vector(1,0){140}}
\put(147,10){\vector(0,1){80}}
\put(135,90){$\alpha$}
\put(270,56){$r$}
\put(196,39){$r_{2,3}$}
\put(157,55){$r_1$}

\put(284,42){\vector(1,0){140}}
\put(287,10){\vector(0,1){80}}
\put(275,90){$\alpha$}
\put(410,50){$r$}
\put(336,35){$r_{1,2,3}$}

\put(52,0){$(a)$}
\put(192,0){$(b)$}
\put(332,0){$(c)$}

\includegraphics[height=3cm]{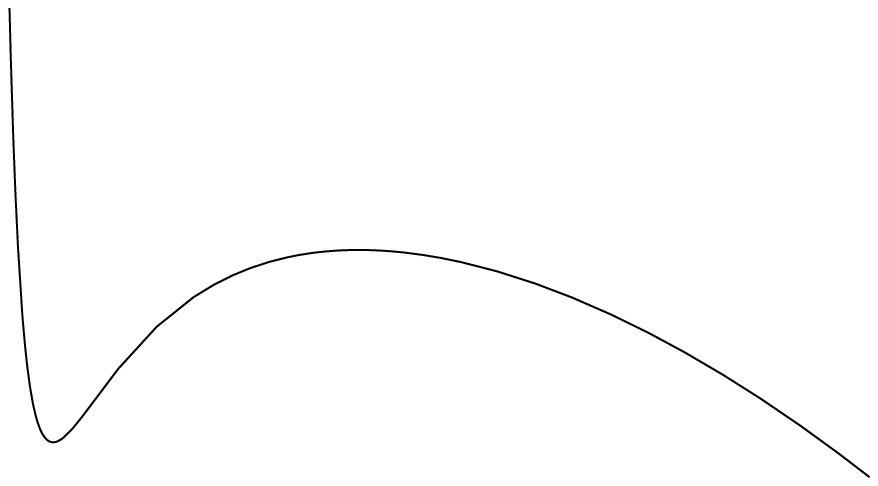}
\includegraphics[height=3cm]{12.eps}
\includegraphics[height=3cm]{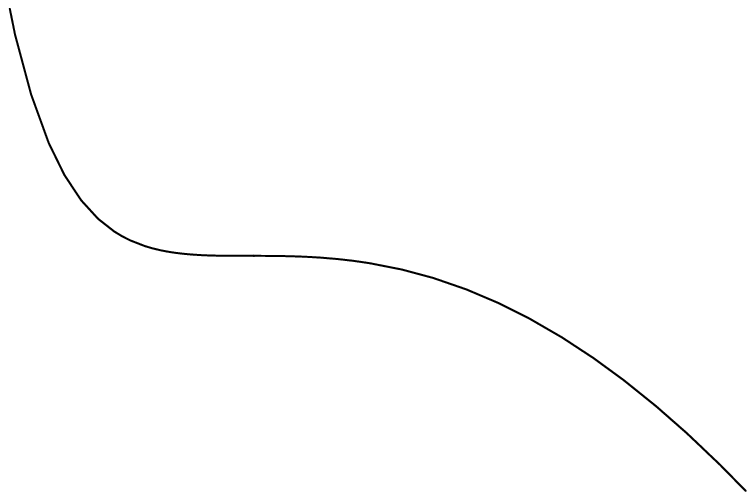}
\end{picture}
\caption{Three types of extremal RN-dS black holes.}\label{fg2}
\end{figure}

A RN-dS black hole is called extremal if two or three horizons coincide. The three types of extremal horizons are illustrated in \fig{fg2}. Using the method suggested by \cite{OBZ2}, it is  easy to check that when
\begin{equation}
M=\frac{\sqrt{6}}{9}l, Q=\frac{\sqrt{3}}{6}l
\end{equation}
is satisfied, the three horizons coincide at
\begin{equation}
r_{1,2,3}=\frac{l}{\sqrt{6}}.
\end{equation}

\section{collision energy in the center-of-mass frame}\label{s2}
In this section, we shall study the collision of two particles in the RN-dS spacetime. Suppose that the two particles have the same mass $m$ and different charges $q_1$ and $q_2$. In the rest of the paper, we shall focus on collisions near the cosmological horizon which is located at $r=r_3$. Collisions near the black hole horizon  $r=r_2$ is essentially the same as the case in the absence of the cosmological constant which has been studied by Zaslavskii \cite{OBZ}.  Suppose the two particles collide in the region $r>r_3$. The center-of-mass energy is given by \cite{MJS}
\begin{equation}\label{Ecm}
E_{c.m.}=m\sqrt{2}\sqrt{1-g_{ab}u^a_1u^b_2},
\end{equation}
 where
\begin{equation}
u^a=\frac{dx^{\mu}}{d\tau}\left(\frac{\partial}{\partial{x^{\mu}}}\right)^a
\end{equation}
is the four-velocity of particle at the collision point. The motion of a charged particle is determined by the following Lagrangian \cite{ruffini}
\bean
\mathscr{L}=\frac{m}{2} g_{\mu\nu}\dot x^\mu \dot x^\nu+q A_\mu\dot x^\mu. \label{gl}
\eean
Here $\dot x^\mu\equiv dx^\mu/d\tau$ and $A_\mu$ are the components of the electromagnetic  4-potential. Substituting the RN-dS metric \meq{ds2} into
\eq{gl}, we find \cite{S.Cha},
\begin{equation}
\mathscr{L}=\frac{m}{2}(-\alpha\dot{t}^2+\frac{1}{\alpha}\dot{r}^2+r^2\dot{\theta}^2+r^2\sin^2\theta\dot{\varphi}^2)-\frac{qQ}{r}\dot{t}.
\end{equation}
The Euler-Lagrange equation then leads to the constant of the motion
\begin{equation}
E=m\alpha\dot{t}+\frac{qQ}{r} \,. \label{td}
\end{equation}
For our purposes, we shall confine our discussion to radial motions, i.e., $\dot\theta=\dot\phi=0$. Thus, the normalization condition  $g_{\mu\nu}u^{\mu}u^{\nu}=-1$ yields
\begin{equation}\label{dr}
\dot{r}=-\frac{1}{m}\sqrt{\left(E-\frac{qQ}{r}\right)^2-m^2\alpha}.
\end{equation}
Here we have chosen the minus sign because $-\ppa{r}{a}$ is a future-directed timelike vector outside the cosmological  horizon.

By direct substitution of \eqs{td} and \meq{dr} in  \eq{Ecm}, we find
\begin{equation}\label{Ecm1}
E^2_{c.m}=2m^2\left[1+\frac{(E_1-\frac{q_1Q}{r})(E_2-\frac{q_2Q}{r})
-\sqrt{(E_1-\frac{q_1Q}{r})^2-m^2\alpha}\sqrt{(E_2-\frac{q_2Q}{r})^2-m^2\alpha}}{m^2\alpha}\right].
\end{equation}
Our purpose is to find conditions for a possible infinite $\ecm$. \eq{Ecm1} suggests that an infinite $\ecm$ can occur only at a horizon where $\alpha=0$. We shall focus on the cosmic horizon located at $r=r_3$. To see if $\ecm$ diverges at the horizon, we need to calculate the limiting value of the numerator of \eq{Ecm1}. The lowest order of the numerator can be obtained by taking $\alpha=0$ and $r=r_H$, which gives
\bean
\left(E_1-\frac{q_1Q}{r_H}\right)\left(E_2-\frac{q_2Q}{r_H}\right)
-\sqrt{\left(E_1-\frac{q_1Q}{r_H}\right)^2}\sqrt{\left(E_2-\frac{q_2Q}{r_H}\right)^2}. \label{num}
\eean
To simplify this formula, we need to know the sign of $E_i-q_iQ/r_H$, where $i=1,2$. According to \eq{td}, $E_i-q_iQ/r_H$  is proportional to  $\dot t$. Note that $\ppa{t}{a}$ is spacelike in the region $r>r_H$. Thus the sign of $\dot t$ corresponds to an ingoing mode or outgoing mode. If the two particles could take different sign at the horizon, it would mean that one particle falls toward the horizon and the other one escapes from the horizon. This is not possible because even a photon cannot escape from the horizon \footnote{The same argument has been used in our paper \cite{gsccz} for the inner horizon of a Kerr black hole.}. Thus, $E_i-q_iQ/r_H$ must have the same sign and then \eq{num} vanishes. The vanishing of \eq{num} is important for our following analysis. Otherwise,  $\ecm$ would be generically divergent without requiring the particle to possess a critical charge.  By expanding the numerator of \eq{Ecm1} around $\alpha=0$, we obtain the following limit
\begin{equation}\label{Ecm2}
\frac{E^2_{c.m}}{2m^2}=1+\frac{1}{2}\left(\frac{q_2-\frac{E_2r_H}{Q}}{q_1-\frac{E_1r_H}{Q}}+\frac{q_1-\frac{E_1r_H}{Q}}{q_2-\frac{E_2r_H}{Q}}\right).
\end{equation}
We see that $\ecm$ blows up at the horizon if one of the particle takes the critical charge
\bean
q_c=\frac{E r\!_H}{Q} \label{qc}
\eean
 and the other particle takes any different value of charge.

Since the divergence of energy occurs at the horizon $r=\rh=r_3$, we need to check if the particle can actually reach the horizon from infinity. This requires that the square root of \eq{dr} must be positive in a vicinity of the horizon. Note that $\alpha<0$ in the region $r>r_3$, it follows immediately that $\dot r^2>0$ outside the cosmological horizon. Thus, with the critical charge, the particle can fall all the way from infinity to the horizon. The above argument does not require any fine tuning on the parameters $Q$, $M$ and $\Lambda$. Thus, an infinite $\ecm$ is attainable at the cosmological horizon of a generic RN-dS black hole.

\section{Numerical estimation}\label{s3}
We see from \eq{qc} that the critical charge is proportional to $r_3$, which is a very large quantity. Consequently, it may lead to a very large charge to mass ratio. To estimate the ratio, we choose $E$ in \eq{qc} to be the proper mass $m$ of the particle. Thus, the required charge to mass ratio in SI units reads
\bean
q/m=\frac{4\pi\epsilon_0 r_3 c^2}{Q}. \label{cmr}
\eean
To estimate $r_3$, write down $\alpha$ in SI units
\bean
\alpha=1-\frac{2GM}{c^2r}+\frac{GQ^2}{4\pi\varepsilon_0c^4r^2}-\frac{G\Lambda r^2}{3c^2}.
\eean
We take $M=2\times 10^{32} kg$, which is about 100 times the solar mass. In geometrized units, the maximum charge of a black hole is $Q=M$. So in SI units, the maximum charge is given by
\bean
Q\sim\sqrt{\frac{9\pi \epsilon_0 G}{2}}M\sim 1.8\times 10^{22} \ Coulomb.
\eean
Based on supernovae observations, the value of the cosmological constant is about
\bean
\Lambda\sim 8\pi\rho_{vac}\times 10^{-120}
\eean
where $\rho_{vac}\sim 10^{94} kg\cdot m^{-3}$ is the vacuum energy density.
Then $r_3$ can be solved as $r_3\sim 10^{26}m$ and the right-hand side of \eq{cmr} is found to be
\bean
q/m\sim 5\times 10^{10} \ coul/kg.
\eean
Note that the charge to mass ratio for an electron is
\bean
e/m_e\sim 1.8\times 10^{11} \ coul/kg
\eean
which is comparable to the required charge to mass ratio.  However, the above estimation is too optimistic because we have chosen the maximum charge $Q$ for the black hole. Such a value only exists in theory. According to \cite{wald},   a real astrophysical body usually has negligible charge which satisfies  $Q/M<10^{-18}$ in geometrized units.  This means that the charge to mass ratio required for a particle is at least $10^{18}$ times larger than that of an election. Such particles are unlikely to exist. In order to find a smaller charge to mass ratio, we have to consider other configurations.  \eq{cmr} suggests that $q/m$ can be decreased by increasing $Q$ or decreasing $r_3$. Since $r_3\sim l=\sqrt{\frac{3}{\Lambda}}$ is almost a constant, the only choice left is to make $Q$ as large as possible.  According to the recent observation and analysis\cite{mass}\cite{mass2}, the most massive black hole may have $10^{10}$ solar masses, i.e., $M\sim 2\times 10^{40} kg$. By taking $Q=10^{-18}M$, we find that $q/m\sim 10^{20}$, still many orders of magnitude larger than $e/m_e$. Thus, the current astrophysical evidences suggest that  the required critical charge is unlikely to be attained by fundamental particles.

\section{conclusions} \label{conc}
We have studied the BSW accelerating effect for a RN-dS black hole. The presence of the cosmological constant makes infinite CM energy possible at the cosmological horizon. Unlike the RN case, the divergence of energy does not require the black hole to be extremal. On the other hand, one of the particles must possess a critical charge. Simple numerical estimation shows that a reasonable charge to mass ratio is possible only when the black hole is nearly extremal, i.e., $Q\sim M$. However, in the real world any astrophysical body may not have a charge to mass ratio of greater than $10^{-18}$. By using this limit, we find that even for the most massive black hole,  the required charge to mass ratio would be much higher than that of electron. Therefore, even in the presence of cosmological constant, a collision with arbitrarily high energy is unlikely to occur due to limitations on the maximal charge to mass ratios of astrophysical bodies and elementary particles.

\section{Acknowledgements}
This research is supported by NSFC grants 10605006, 10975016 and by "the Fundamental Research Funds for the Central Universities".

\end{document}